\documentclass[sigconf]{acmart}
\usepackage{subfigure}
\usepackage{diagbox}
\usepackage{multirow}
\usepackage{amsmath,bm}

\AtBeginDocument{%
  \providecommand\BibTeX{{%
    \normalfont B\kern-0.5em{\scshape i\kern-0.25em b}\kern-0.8em\TeX}}}
\copyrightyear{2021}
\acmYear{2021}
\setcopyright{acmcopyright}\acmConference[CIKM '21]{Proceedings of the 30th ACM International Conference on Information and Knowledge Management}{November 1--5, 2021}{Virtual Event, QLD, Australia}
\acmBooktitle{Proceedings of the 30th ACM International Conference on Information and Knowledge Management (CIKM '21), November 1--5, 2021, Virtual Event, QLD, Australia}
\acmPrice{15.00}
\acmDOI{10.1145/3459637.3482181}
\acmISBN{978-1-4503-8446-9/21/11}

\begin{document}
\fancyhead{}

\title{SIFN: A Sentiment-aware Interactive Fusion Network for Review-based Item Recommendation}

{\author{Kai Zhang$^{1}$, Hao Qian$^3$, Qi Liu$^{1,2,*}$, Zhiqiang Zhang$^{3}$, Jun Zhou$^{3}$, Jianhui Ma$^{1}$, Enhong Chen$^{1,2}$}}
\thanks{* Corresponding Author.}
\affiliation{$^{1}$Anhui Province Key Lab of Big Data Analysis and Application, School of Data Science, \\University of Science and Technology of China; $^{3}$Ant Group, Hangzhou, China}
\affiliation{$^{2}$School of Computer Science and Technology, University of Science and Technology of China}
\affiliation{kkzhang0808@mail.ustc.edu.cn, \{qiliuql, jianhui, cheneh\}@ustc.edu.cn\\
\{qianhao.qh, lingyao.zzq, jun.zhoujun\}@antgroup.com}

\begin{abstract}
Recent studies in recommender systems have managed to achieve significantly improved performance by leveraging reviews for rating prediction. However, despite being extensively studied, these methods still suffer from some limitations. First, previous studies either encode the document or extract latent sentiment via neural networks, which are difficult to interpret the sentiment of reviewers intuitively. Second, they neglect the personalized interaction of reviews with user/item, i.e., each review has different contributions when modeling the sentiment preference of user/item. 

To remedy these issues, we propose a \underline{\textbf{S}}entiment-aware \underline{\textbf{I}}nteractive \underline{\textbf{F}}usion \underline{\textbf{N}}etwork ({\textbf{SIFN}}) for review-based item recommendation. Specifically, we first encode user/item reviews via BERT and propose a light-weighted sentiment learner to extract semantic features of each review. Then, we propose a sentiment prediction task that guides the sentiment learner to extract sentiment-aware features via explicit sentiment labels. 
Finally, we design a rating prediction task that contains a rating learner with an interactive and fusion module to fuse the identity (i.e., user and item ID) and each review representation so that various interactive features can synergistically influence the final rating score. Experimental results on five real-world datasets demonstrate that the proposed model is superior to state-of-the-art models.
\end{abstract}

%


\maketitle
{\fontsize{8pt}{8pt} \selectfont
	\textbf{ACM Reference Format:}\\
	Kai Zhang, Hao Qian, Qi Liu, Zhiqiang Zhang, Jun Zhou, Jianhui Ma, Enhong Chen. 2021. SIFN: A Sentiment-aware Interactive Fusion Network for Review-based Item Recommendation. In \textit{Proceedings of the 30th ACM International Conference on Information and Knowledge Management (CIKM ’21), November 1–5, 2021, Virtual Event, QLD, Australia}. ACM, New York, NY, USA, 5 pages. https://doi.org/10.1145/3459637.3482181
}

\section{Introduction}
\label{intro}
Recommender Systems (RS) have been widely adopted by many platforms, e.g. Amazon and Yelp, thanks to their capability of filtering information based on the user interests~\cite{hou2019explainable,zhang2021multi}. In these platforms, user reviews of items contain rich semantic information. Therefore, to better estimate user's rating, review-based recommendation has gained wide attention in both academia~\cite{bao2014topicmf} and industry~\cite{seo2017interpretable,li2019capsule}.

In the literature, there are many efforts for this recommendation problem, especially on the rating prediction task. Among them, the most representative methods are based on Matrix Factorization (MF) model, which decomposes the user-item rating matrix into two matrices corresponding to user and item features. However, these methods represent user and item information only based on ratings, which suffer from sparsity issue. To mitigate this problem, many RS have been proposed to exploit the semantic information from user reviews
~\cite{bao2014topicmf,tan2016rating,zheng2017joint,tay2018multi,chen2018neural}. Among these review-based methods, the earlier works adopt MF to extract the semantics from user/item reviews, such as PMF\cite{mnih2007probabilistic} and ConvMF+~\cite{kim2016convolutional}. Recently, the focus of research shifts to learn latent features from reviews via neural network methods. Among them, DeepCoNN~\cite{zheng2017joint} applies convolution to process the reviews to learn user and item representations. D-Attn~\cite{seo2017interpretable} leverages global and local attention to select decisive words in the review documents for rating prediction. NARRE~\cite{chen2018neural} is a review retrieval model that adopts an attention mechanism to select appropriate reviews for items. Moreover, CARP~\cite{li2019capsule} utilizes a sentiment capsule network to estimate user-item ratings and provide interpretability with a fine-grained manner.

Although these works achieved significant performance improvement in the review-based recommendation, they still suffer from intrinsic issues. First, most previous works directly encode reviews to extract implicit semantics while largely ignoring the explicit sentiment polarity of reviews~\cite{seo2017interpretable,tay2018multi}, which carry the user attitudes and preferences (i.e., which kinds of item user may like or dislike). Therefore, implicitly mining the semantic information of reviews may lead to sub-optimal prediction because the reviews' sentiment label has not been applied to the training process~\cite{zhang2014explicit,he2015trirank,ren2017social,tay2018multi}. Second, in the rating prediction scenario, reviews' sentiment polarity plays distinct roles in representing the overall sentiment preference of the user/item. For instance, the impact of a negative review may be far more significant than a positive review for the user/item's semantic preference representation. Therefore, it is necessary to weigh and fuse the interactions between each review and user/item features for the final rating prediction. 

\begin{figure*}
	\centering
	\vspace{-0.1cm}
	\includegraphics[width=1.95\columnwidth]{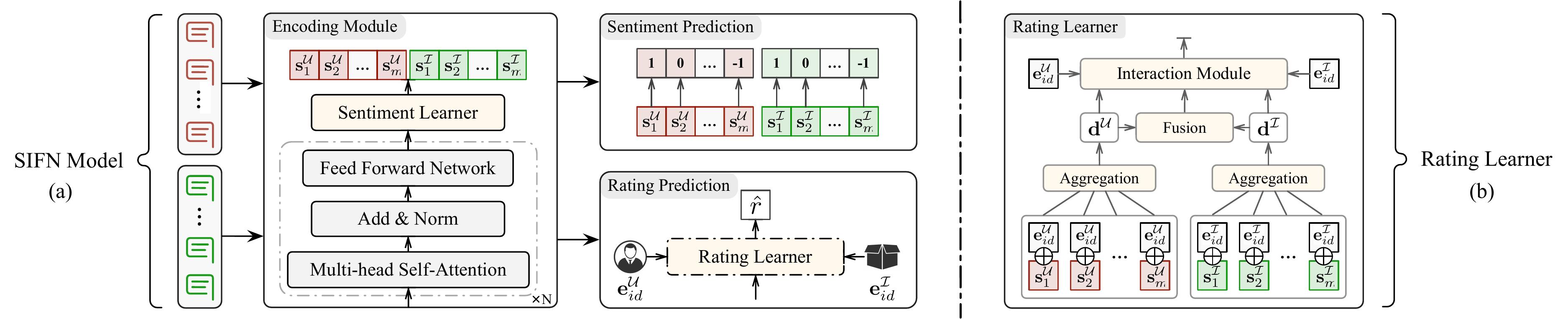}
	\vspace{-0.15cm}
	\caption{(a) The overall architecture of SIFN model; (b) the rating learner which can exploit interactive fusion knowledge.
	}
	\vspace{-0.1cm}
	\label{fig:model}
\end{figure*}

To address these aforementioned issues, we propose a {\textbf{S}}entiment-aware {\textbf{I}}nteractive {\textbf{F}}usion {\textbf{N}}etwork ({\textbf{SIFN}}), which includes a sentiment prediction task to exploit the sentiment polarity from each review, and a rating prediction task to estimate user rating of items. Specifically, we first utilize the pre-trained BERT to encode text reviews and introduce a light-weighted sentiment learner to mine the sentiment-aware features. Then, we design a sentiment prediction task that explicitly extracts important sentiment features in each review for user/item. Finally, in the rating prediction task, we develop a rating leaner which contains two novel operations (i.e., interactive \& fusion) to fuse the identity representation of user/item and review representation so that various interactive features can synergistically affect the final rating prediction. The main contributions of this paper are as follows:  
1) We highlight the explicit sentiment polarity in each review, and focus on modeling the multiple feature interactions between each review and user/item.
2) We propose a novel {\textbf{S}}entiment-aware {\textbf{I}}nteractive {\textbf{F}}usion {\textbf{N}}etwork ({\textbf{SIFN}}) model with two main components, Sentiment Leaner and Rating Learner. 
3) We conduct extensive experiments on five datasets so that the results demonstrate the effectiveness of the proposed method.

\section{The Proposed Model}
Figure~\ref{fig:model} illustrates the overall architecture of SIFN, which mainly consists of three components. The details of each component will be explained in the following sections.

\subsection{The Encoding Module}
In this section, we present how to extract deep semantic representations of reviews progressively. Since the semantic learning process of reviews for user and item are quite similar, we mainly introduce user reviews' semantic representation for space saving. 

\vspace{0.035cm}
\textbf{BERT Encoding.} Given a user review $\mathbf{x}^\mathcal{U} = (w^\mathcal{U}_1, w^\mathcal{U}_2, ..., w^\mathcal{U}_l) \in \mathbb{R}^{l}$ which is formed by merging the total $l$ words in a review written by a user. For word embedding, we adopt the pre-trained BERT~\cite{devlin2018bert} to encode each word and use the last hidden state of the pre-trained BERT for word representations~\cite{zhang2021eatn}. The formulation is as follows:
\begin{equation}
	\begin{aligned} 
	\mathbf{R}^\mathcal{U} = (\mathbf{e}^\mathcal{U}_1, \mathbf{e}^\mathcal{U}_2, ..., \mathbf{e}^\mathcal{U}_l) = {\rm BERT}(w^\mathcal{U}_1, w^\mathcal{U}_2, ..., w^\mathcal{U}_l),
	 \end{aligned}
\end{equation}
where $\mathbf{R}^\mathcal{U} \in \mathbb{R}^{l \times k}$ is the embedding of a user review, $\mathbf{e}^\mathcal{U}_l \in \mathbb{R}^{k}$ is the embedding of the $l$-th word, where $k$ is the word embedding size. Similarly, an item review can be mapped as $\mathbf{R}^\mathcal{I}=(\bm{e^\mathcal{I}_1, e^\mathcal{I}_2, ..., e^\mathcal{I}_l)} \in \mathbb{R}^{l \times k}$. With the powerful pre-trained models, the general semantic knowledge of the text reviews can be fully extracted.

\vspace{0.035cm}
\textbf{Sentiment Learner.} As each word in a review carries distinct semantic and sentimental information, which is crucial to estimate user rating and sentiment polarity. Therefore, to better suit the downstream tasks, we devise a sentiment learner to adaptively underline the informative words. Specifically, we employ the attention mechanism, in which the attention score of $i$-th word is calculated as:
\begin{equation}
\alpha_{i}=\frac{\exp (\tanh ({\mathbf{W}_a \mathbf{e}^\mathcal{U}_i} +\mathbf{b}_a))}{\sum_{i^{\prime}=1}^{l} \exp (\tanh ( \mathbf{W}_{a} \mathbf{e}^\mathcal{U}_{i^{\prime}} +\mathbf{b}_a))}.
\end{equation}

$\mathbf{W}_a$ and $\mathbf{b}_a$ are weight matrix and bias, respectively. $\alpha_i$ is attention score of $i$-th word in a review. Therefore, words with smaller attention score are less important. Further, with a weighted sum pooling over the words in a review, the effects of uninformative words are diminished and the semantic representation of the user review $\mathbf{s}^\mathcal{U} \in \mathbb{R}^{k}$ is calculated as:
\begin{equation}
\mathbf{s}^\mathcal{U}=\sum_{i=1}^{l} \alpha_{i} \mathbf{e}^\mathcal{U}_i, 
\end{equation}

Through the encoding module, we obtain each review's representation of the user, i.e., $\mathbf{s}^\mathcal{U}$, via aggregating feature vectors of the words. Similarly, we use the same method to generate each review's representation, i.e., $\mathbf{s}^\mathcal{I} \in \mathbb{R}^{k}$, for the item. 

\subsection{Sentiment Prediction}
Since the sentiment polarity of review carries the intrinsic user preference toward item, we design a sentiment prediction task for each review to learn sentiment features explicitly (e.g., $\mathbf{s}^\mathcal{U}$ and $\mathbf{s}^\mathcal{I}$). Note that, the ground-truth is obtained by converting the user-item rating to sentiment label with a threshold of 3, i.e., we category them into positive (i.e., higher than 3 stars), negative (i.e., lower than 3 stars) or neural (i.e., equal to 3 stars). 

Hereby, we utilize a cross entropy loss function to estimate the sentiment polarity of each user and item review as follows:
\begin{equation}
	L_{s}= -\frac{1}{|O|} \sum_{(\mathcal{U}, \mathcal{I}) \in O} \{\frac{1}{m} \sum_{j=1}^{m}(\sum_{t=1}^{C} y_{j}^{t} \log (\hat{y}_{j}^{t})) \},
\end{equation}
where $\hat{y}_j^t= softmax( \mathbf{s}^\mathcal{U}_j)$ is the predicted sentiment polarity of the $j$-th user review. Similarly, we can get sentiment prediction of item review $\hat{y}_j^t=softmax( \mathbf{s}^\mathcal{I}_j)$. $O$ denotes the set of observed user-item rating pairs. $C$ is the number of sentiment polarity categories and $m$ is the number of reviews for a user/item.

\subsection{Rating Prediction} 
In this subsection, we apply our SIFN model for the rating prediction task. To this end, we propose a rating learner to comprehensively extract the latent interactions among the identity representation of user/item and review representation, as shown in figure~\ref{fig:model} (b).

To begin with, there is user ID feature, which is encoded into low-dimensional vector denoted as  $\mathbf{e}^{\mathcal{U}}_{id} \in \mathbb{R}^{k}$ through:
\begin{equation}
\mathbf{e}^{\mathcal{U}}_{id} = \mathbf{W}^{\mathcal{U}}_{id} \cdot \mathbf{x}^\mathcal{U}_{id} ,
\end{equation}
where $\mathbf{W}^{\mathcal{U}}_{id} \in \mathbb{R}^{n^\mathcal{U} \times k}$ is weight matrix and $n^\mathcal{U}$ is the total number of user ID features.

\vspace{0.05cm}
\textbf{Rating Learner.} As previous work states~\cite{chen2018neural}, different reviews do not contribute equally to the preference of user/item. Besides, user also carries distinct style when writing reviews. For example, an optimistic user is more likely to write positive reviews while a cranky user tends to write something even rude. Therefore, in order to better model the user/item's personalized preference in different reviews, it is indispensable to incorporate both text review and user/item features. Hereby, we concatenate them as follows:
\begin{equation}
	\vspace{-0.05cm}
	\begin{aligned} 
	\mathbf{o}^{\mathcal{U}}_j = \mathbf{W}_o \cdot {\rm concat} (\mathbf{s}^\mathcal{U}_j;\mathbf{e}^{\mathcal{U}}_{id}),
	 \end{aligned}
\end{equation}
where $\mathbf{W}_o \in \mathbb{R}^{k \times 2k}$ is a projection weight matrix.

\vspace{0.05cm}
\emph{\textbf{1) Aggregation.}} Then, we utilize the user-aware review representation $\mathbf{o}^{\mathcal{U}}_j \in \mathbb{R}^{k}$ to adaptively select important reviews with attention function which is formulated as:
\begin{equation}
\mathbf{d}^{\mathcal{U}} =\sum_{j=1}^{m}{\frac{\exp \left(\tanh \left({\mathbf{W}_o \mathbf{o}^{\mathcal{U}}_j } +\mathbf{b}_o\right)\right)}{\sum_{j^{\prime}=1}^{m} \exp \left(\tanh \left( \mathbf{W}_o \mathbf{o}^{\mathcal{U}}_{j^{\prime}} +\mathbf{b}_o\right)\right)}} \mathbf{o}^{\mathcal{U}}_j,
\end{equation}
where $\mathbf{d}^{\mathcal{U}} \in \mathbb{R}^{k}$ is the user review representation with aggregated information from all user reviews under the influence of user features. Similarly, we can get the embedding vector of item ID denoted as $\mathbf{e}^{\mathcal{I}}_{id} \in \mathbb{R}^{k}$ and the item review representation $\mathbf{d}^{\mathcal{I}} \in \mathbb{R}^{k}$.

\vspace{0.05cm}
\emph{\textbf{2) Fusion Network.}} So far, the user/item reviews representation (i.e., $\mathbf{o}^{\mathcal{U}}_j$ and $\mathbf{o}^{\mathcal{I}}_j$), that integrates the features of identity and reviews, are learned from separate hierarchical representation modules without interactions. Since we aim at learning the user-item rating, it is necessary to extract the mutual influence among them. However, as user and item features may carry different characteristics in different spaces, we design a novel fusion network that projects features into a shared hidden space before learning feature interactions as:
\begin{equation}
	\mathbf{f} = \mathbf{d}^{\mathcal{U}} \mathbf{W}_f \mathbf{d}^{\mathcal{I}} ,
\end{equation}
where $\mathbf{f} \in \mathbb{R}^{k}$ is the fusion representation, $\mathbf{W}_f \in \mathbb{R}^{k \times k}$ is the weight matrix that maps user and item features to the same latent space.

\vspace{0.05cm}
\emph{\textbf{3) Interactive Network.}} In addition, to underline the contribution of the different extracted representations (e.g., $\mathbf{d}^{\mathcal{U}}$, $\mathbf{d}^{\mathcal{I}}$ and $\mathbf{f}$), we further design an interactive network so that the user-item rating estimation is fully explored as follows:
\begin{equation}
\mathbf{p} = \left( \mathbf{d}^{\mathcal{U}}+\mathbf{e}^{\mathcal{U}}_{id}\right) \odot\left(\mathbf{d}^{\mathcal{I}}+\mathbf{e}^{\mathcal{I}}_{id}\right) + \mathbf{W} \mathbf{f} + \mathbf{b},
\end{equation}
where $\mathbf{W}$ and $\mathbf{b}$ are weight matrix and bias. $\odot$ denotes the element-wise product of vectors. $\mathbf{p}\in \mathbb{R}^{k}$ is the user-item preference representation with aggregated interactions from reviews and identity.

Through the above learning processes, the predicted rating of user toward item $\hat{r}$ is obtained by a linear projection and  the regression loss function for rating prediction is as:
\begin{equation}
	\hat{r} = \mathbf{w}_r \mathbf{p} + b^\mathcal{U} +  b^\mathcal{I};\ \  L_{r}=\frac{1}{|O|} \sum_{O} \left(r-\hat{r}\right)^{2},
\end{equation}
where $\mathbf{w}_r$ is the weight parameter. $b^\mathcal{U}$ and $b^\mathcal{I}$ are biases for user and item, respectively. $r$ is the ground truth user-item rating.

Finally, we utilize the joint learning process to optimize both objectives with a hyper-parameters $\lambda$ as: 
\begin{equation}
L =  L_r + \lambda \cdot L_s.
\end{equation}

\section{Experiment}
\subsection{Dataset}
We choose five Amazon datasets (i.e., Music Instruments, Office Products, Digital Music, Tools, Video Games) from ~\cite{li2019capsule} to conduct our experiments. We preprocessed it to ensure that all users and items have at least one rating five reviews in our experiment. The dataset consists of numerous data samples on which we follow a randomized 80:10:10 train/test/validation split. For concrete information, please refer to Table~\ref{tab:data}.

\subsection{Baseline methods}
To show the performance of our proposed model, we compare SIFN with the state-of-the-art (SOTA) models. The benchmarks are:
\begin{itemize}
	\item \textbf{MF-based}: PMF~\cite{mnih2007probabilistic} models the latent factors by introducing Gaussian distribution. ConvMF+~\cite{kim2016convolutional} incorporates convolutional neural network into Matrix Factorization.
	\item \textbf{Neural-based}: DeepCoNN~\cite{zheng2017joint}, D-Attn~\cite{seo2017interpretable}, NARRE~\cite{chen2018neural}, CARP~\cite{li2019capsule}. These methods have been introduced in section~\ref{intro}.
\end{itemize}

\begin{table}
    \centering
    \caption{Statistics of the Amazon datasets.} 
    \vspace{-0.15cm}
    \label{tab:data} 
    \begin{tabular}{p{2.35cm}|p{1.0cm}<{\centering}|p{1.0cm}<{\centering}|p{1.2cm}<{\centering}|p{1.1cm}<{\centering}}
            \toprule
            \specialrule{0em}{2pt}{0pt}
            Dataset&\#\ Users&\#\ Items&\#\ Ratings& Density\\
            \midrule
            Music Instruments & 1,429& 900& 10,261& 0.798\%\\
            \specialrule{0em}{1pt}{0pt}
            Office Products & 4,905& 2,420& 53,228& 0.448\%\\
            \specialrule{0em}{1pt}{0pt}
            Digital Music & 5,540 & 3,568& 64,664& 0.327\%\\
            \specialrule{0em}{1pt}{0pt}
            Tools & 16,638 & 10,217& 134,345& 0.079\%\\
            \specialrule{0em}{1pt}{0pt}
            Video Games & 24,303& 10,672& 213,577& 0.089\% \\
            \bottomrule
    \end{tabular}
    \vspace{-0.3cm}
\end{table}

In our SIFN model, we use BERT encoding from Hugging Face~\footnote{https://huggingface.co/transformers/}. User and item embedding size is set to 16. The batch size is 100, learning rate is 0.001, dropout rate is 0.2, and $\lambda$ is tuned amongst [0.1,1,10]. For baseline methods, we follow the hyper-parameter configurations in their papers. Following previous works~\cite{seo2017interpretable,li2019capsule}, we utilize Mean Squared Error (MSE) as the evaluation metric and select Adam as optimizer for all models.

\begin{table*} 
    \caption{Experimental results in Amazon datasets. Percentage in ( ) denotes the relative improvement of SIFN over the baseline method, which is achieved with paired t-tests at the significance level of 0.01. We underline the best performed baseline.} 
    \label{tab:result} 
    \centering
    \vspace{-0.35cm}
    \subtable{
        \begin{tabular}{p{2.2cm}|p{2.4cm}<{\centering}|p{2.2cm}<{\centering}|p{2.2cm}<{\centering}|p{2.1cm}<{\centering}|p{2.1cm}<{\centering}|p{1.8cm}<{\centering}}
            \toprule
            \specialrule{0em}{1pt}{0pt}
            \specialrule{0em}{2pt}{0pt}
             {Methods } &Music Instruments & Office Products & Digital Music & Tools & Video Games & {Average}\\
            \midrule 
            \ \ PMF ~\cite{mnih2007probabilistic}
            & 1.398\small{\emph{(+45.7\%)}}
            &1.092\small{\emph{(+35.7\%)}}
            &1.206\small{\emph{(+33.7\%)}}
            &1.566\small{\emph{(+39.3}\%)}
            &1.672\small{\emph{(+37.4\%)}}
            &1.386\small{\emph{(+38.6\%)}}
            \\
            \specialrule{0em}{1pt}{0pt}
            \ \ ConvMF+ ~\cite{kim2016convolutional}
            &0.991\small{\emph{(+23.4\%)}}
            &0.960\small{\emph{(+26.9\%)}}
            &1.084\small{\emph{(+26.3\%)}}
            &1.240\small{\emph{(+23.4\%)}}
            &1.449\small{\emph{+27.7\%)}}
            &{1.145\small{\emph{(+25.7\%)}}}
            \\
            \specialrule{0em}{1pt}{0pt}
            \ \ DeepCoNN ~\cite{zheng2017joint}
            & 0.814\small{\emph{(+6.76\%)}}
            &0.860\small{\emph{(+18.4\%)}}
            &1.058\small{\emph{(+24.5\%)}}
            &1.061\small{\emph{(+10.5\%)}}
            &1.145\small{\emph{(+8.56\%)}}
            &{0.988\small{\emph{(+13.9\%)}}}
            \\
            \specialrule{0em}{1pt}{0pt}
            \ \ D-Attn ~\cite{seo2017interpretable}
            &0.982\small{\emph{(+22.7\%)}}
            &0.825\small{\emph{(+14.9\%)}}
            &0.911\small{\emph{(+12.3\%)}}
            &1.043\small{\emph{(+8.92\%)}}
            &1.144\small{\emph{(+8.48\%)}}
            &{0.981\small{\emph{(+13.3\%)}}}
            \\
            \specialrule{0em}{1pt}{0pt}
            \ \ NARRE~\cite{chen2018neural}
            &0.803\small{\emph{(+5.48\%)}}
            &0.848\small{\emph{(+17.2\%)}}
            &0.898\small{\emph{(+11.0\%)}}
            &1.029\small{\emph{(+7.68\%)}}
            &1.129\small{\emph{(+7.26\%)}}
            &{0.941\small{\emph{(+9.6\%)}}}
            \\
            \specialrule{0em}{1pt}{0pt}
            \ \ CARP~\cite{li2019capsule}
            &\underline{0.773}\small{\emph{(+1.81\%)}}
            &\underline{0.719}\small{\emph{(+2.36\%)}}
            &\underline{0.820}\small{\emph{(+2.56\%)}}
            &\underline{0.960}\small{\emph{(+1.04\%)}}
            &\underline{1.084}\small{\emph{(+3.41\%)}}
            &\underline{0.872}\small{\emph{(+2.4\%)}}
            \\
            \midrule
            \specialrule{0em}{1pt}{0pt}
            \midrule
            \ \ \textbf{SIFN} 
            &\textbf{0.759}
            &\textbf{0.702}
            &\textbf{0.799}
            &\textbf{0.950}
            &\textbf{1.047}
            &\textbf{0.851}
            \\
            \bottomrule
        \end{tabular}
        \label{tab:melody_acc}  
    } 
\end{table*}

\begin{figure*}
	\centering
	\vspace{-0.1cm}
	\includegraphics[width=2\columnwidth]{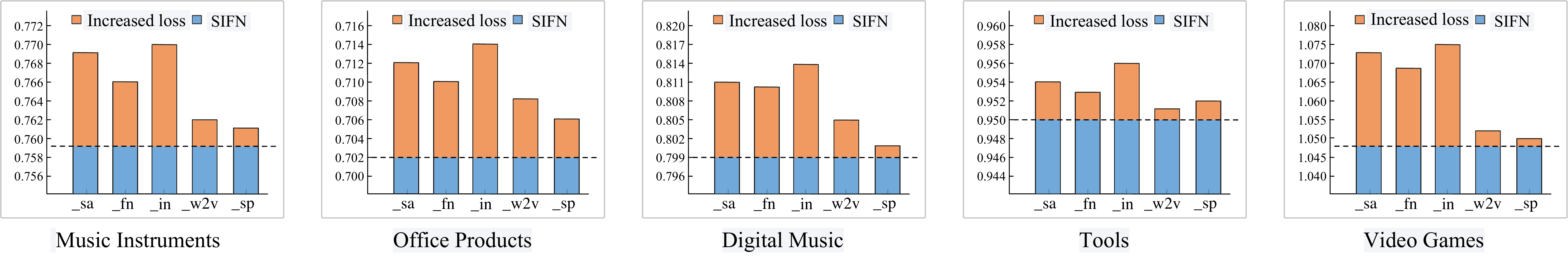}
	\vspace{-0.25cm}
	\caption{Results from variants of SIFN. In each plot, the x-axis denotes the name of variants, in which the prefix SIFN is omitted for space saving. The y-axis denotes the value of evaluation loss. The blue bar represents the loss from SIFN while the stacked orange part denotes increased loss, which indicates a worse performance of variant than SIFN.
	}
	\vspace{-0.15cm}
	\label{fig:ablation}
\end{figure*}

\subsection{Experimental Results}
In this section, we evaluate the performance of SIFN model on Amazon datasets along with detailed comparison of results in Table~\ref{tab:result}. The major results are summarized as follows: (1) MF-based methods (e.g., PMF and ConvMF+) consistently fall behind other methods, which indicates the limitations of Matrix Factorization to learn semantic information from sparse rating dataset. (2) Neural-based methods (e.g., DeepCoNN, D-Attn, NARRE, CARP) outperform MF-based ones by a large margin, which validates the powerful feature extraction capability of neural networks. 
Among them, CARP performs best, which suggests it's helpful to encode viewpoint of user and sentiment aspect of item for rating prediction. (3) Furthermore, our proposed SIFN model still outperforms CARP by 1.81\%$\sim$3.41\%, which demonstrates the superiority of the well designed interactive\& Fusion module that extracts the mutual influence among user and item features and the effectiveness of the sentiment prediction task that attends the sentimental phrases in text reviews.

\subsection{Ablation Studies}
\label{sec:ablation_study}
To explore the impact of each component of SIFN, multiple ablation studies are carried out by removing one sub-module at a time.
\begin{itemize}
	\item \textbf{SIFN\_{sa}}: replaces the sentence attention with a simple average sum pooling over all the reviews.
	\item \textbf{SIFN\_{fn}}: removes the fusion network so that user and item features are disentangled without explicit interactions. 
	\item \textbf{SIFN\_{in}}: replaces the interactive network with commonly used Factorization Machine (FM)~\cite{rendle_fm} to estimate the ratings.
	\item \textbf{SIFN\_{w2v}}: replaces the BERT encoding of text reviews with commonly used pre-trained word embedding GloVe~\cite{pennington-etal-2014-glove}.
	\item \textbf{SIFN\_{sp}}: removes the sentiment prediction task so that the model focuses on user-item rating prediction.
\end{itemize}

We report the results of ablation studies in Figure~\ref{fig:ablation}. Specifically, the performance of \textbf{SIFN\_{sa}} with average pooling drops as it just assumes every review contributes equally for user-item rating, which overlooks the impact of various user/item information. Without the fusion network, the performance of \textbf{SIFN\_{fn}} also declines since the interactions of user and item reviews are not fully exploited. As user and item features carry different characteristics, it is inefficient to perform second-order operations in the same space with FM, which leads to the decrease of the performance of \textbf{SIFN\_{in}}. Not surprisingly, without BERT encoding, the word embedding in \textbf{SIFN\_{w2v}} is incapable of representing deep semantic information of text reviews. Without sentiment prediction in \textbf{SIFN\_{sp}}, there is no supervision towards attending sentiment-aware words in reviews, which are vital for the rating prediction.

\begin{figure}
	\centering
	\vspace{0.1cm}
	\includegraphics[width=0.98\columnwidth]{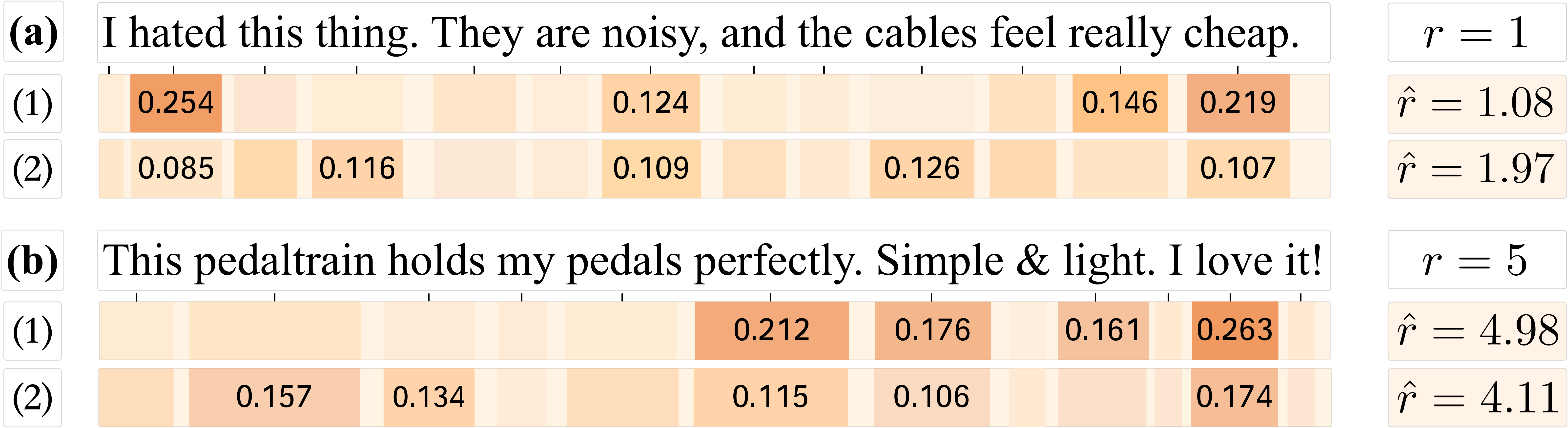}
	\vspace{-0.1cm}
	\caption{Attention visualization of the different words. The word attention value from SIFN and SIFN\_sp are associated with the row (1), row (2) respectively in both examples.
	}
	\label{fig:case}
	\vspace{-0.25cm}
\end{figure}

\subsection{Case Study}
As shown in Figure~\ref{fig:case}, we randomly sample two reviews of a user from the Music Instrument dataset and visualize the attention results to explain the capability of SIFN in extracting sentiment knowledge. Specifically, in review (a), SIFN aligns the sentiment words, e.g., ``hated'' and ``cheap'', with a rating of 1.08, which is consistent with actual value. In contrast, we utilize the attention results from \textbf{SIFN\_{sp}} as described in Section~\ref{sec:ablation_study}, which overlooks these sentiment words thus predicting a relatively higher rating. Similarly, in review (b), SIFN accurately predicts a rating of 4.98 by extracting the sentiment words, e.g., ``love'' and ``perfectly'', while \textbf{SIFN\_{sp}} is incapable of achieving this. These examples demonstrate that SIFN is effective in extracting the latent semantics of reviews and interpret the corresponding sentiment, which is helpful for the final rating prediction.

\section{Conclusions}
In this paper, we proposed a novel Sentiment-aware Interactive Fusion Network (SIFN) model for review-based item recommendation. Specifically, we first employed the encoding module which contains BERT encoding and a sentiment learner to learn sentiment-aware features of each review sentence. Then, we designed a sentiment prediction task to guide the learn process of sentiment features. Finally, we developed a rating learner for the final rating prediction, which contains two novel operations (i.e., interactive \& fusion) to fuse the identity features of user/item and review representation. Extensive experiments on five public datasets demonstrate the effectiveness of our proposed method.

\section{Acknowledgment}
This research was partially supported by grants from the National Natural Science Foundation of China (Grants No. 61922073 and U20A20229). 

\bibliographystyle{ACM-Reference-Format}
\bibliography{sample-sigconf.bib}
\end{document}